\documentclass[review]{elsarticle}

\usepackage{lineno,hyperref}
\usepackage{hyperref}
\modulolinenumbers[5]

\usepackage{graphicx}
\usepackage{url}

\begin{document}

\begin{frontmatter}

\title{Theory Summary 
}

\author{Tetsufumi Hirano\fnref{}}
\address{Department of Physics, Sophia Univeristy, Tokyo 102-8554, Japan}
\ead{hirano@sophia.ac.jp}




\begin{abstract}
In this review, I show a personal overview of theoretical results
shown in the International Conference on the Initial Stages in High-Energy Nuclear Collision, in Illa da Toxa, Galicia, Spain,
Sept.~8-14, 2013.
\end{abstract}


\end{frontmatter}


\section{Introduction}

There are 25 plenary and 33 parallel theory talks in
the International Conference on the Initial Stages of High-Energy Nuclear Collisions (IS2013), 
so I must be very selective and summarize them only from a personal point of view.
Therefore readers are advised to browse slides of the talks which
can be found in the website of the conference \cite{is2013} together with
these proceedings.

\begin{figure}[htbp]
  \centering
  \includegraphics[width=10cm, clip, bb=0 0 720 560]{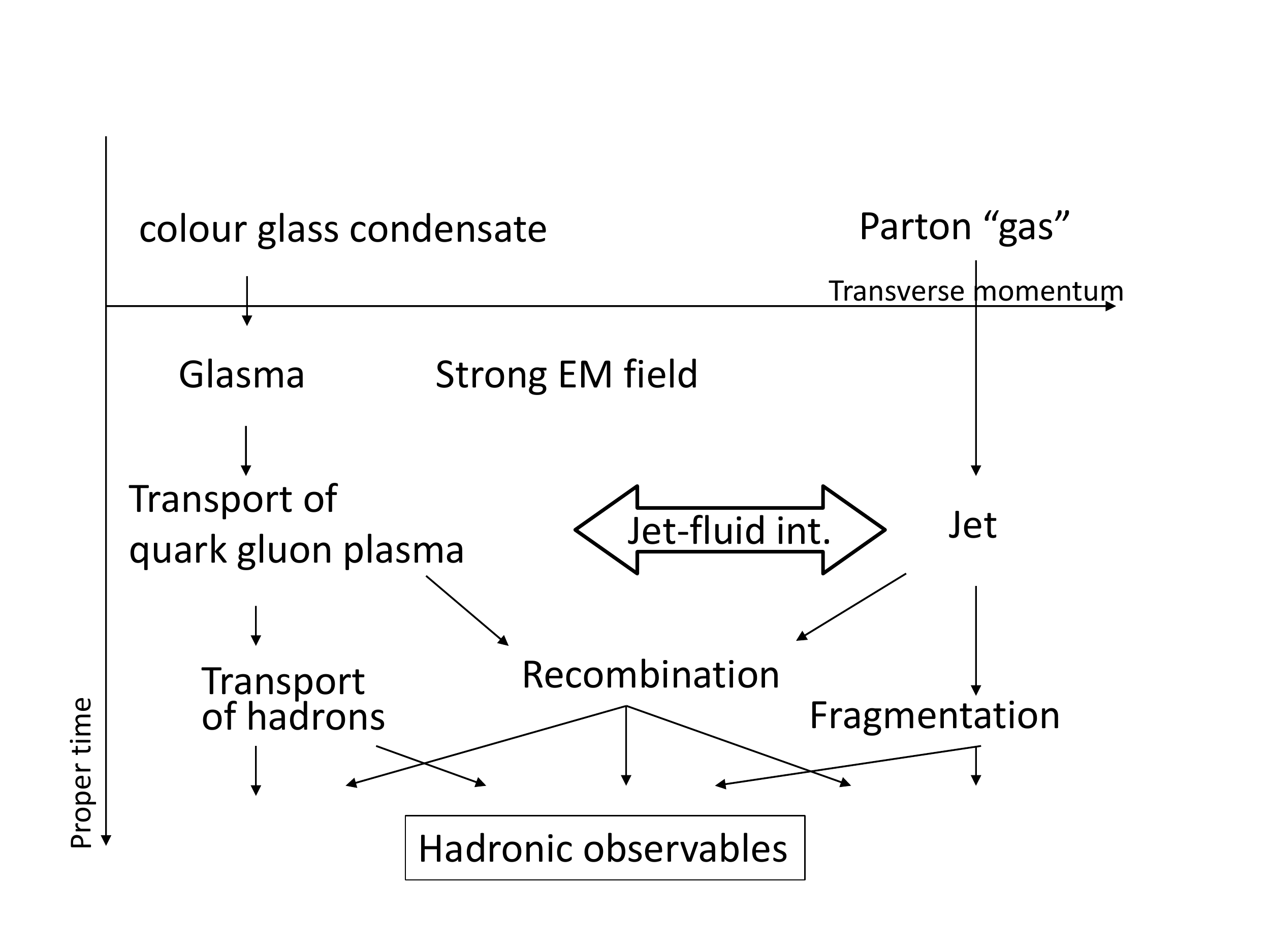}
  \caption{
Topics of this conference and their mutual relations.
\label{fig:crosstalk}
}
\end{figure}

Before going to details about each topic, I 
show two theoretical results which led to motivation for holding this conference.
According to the local organizers,
the aim of the conference is to set a framework of cross-talk among researchers
who conduct a research on initial stages of high energy nuclear collisions
\textit{and} on final hydrodynamic evolutions
along with other topics such as nuclear parton distribution function and thermalization just after
collisions:
The name of the conference apparently indicates
topics covering only the former, which
is not actually true.
All topics shown in Fig.~\ref{fig:crosstalk}  are intimately related with each other: Final observables to be
compared with experimental data originate from convolution of them.

\begin{figure}[htbp]
  \centering
  \includegraphics[width=10cm, clip, bb=0 20 450 300]{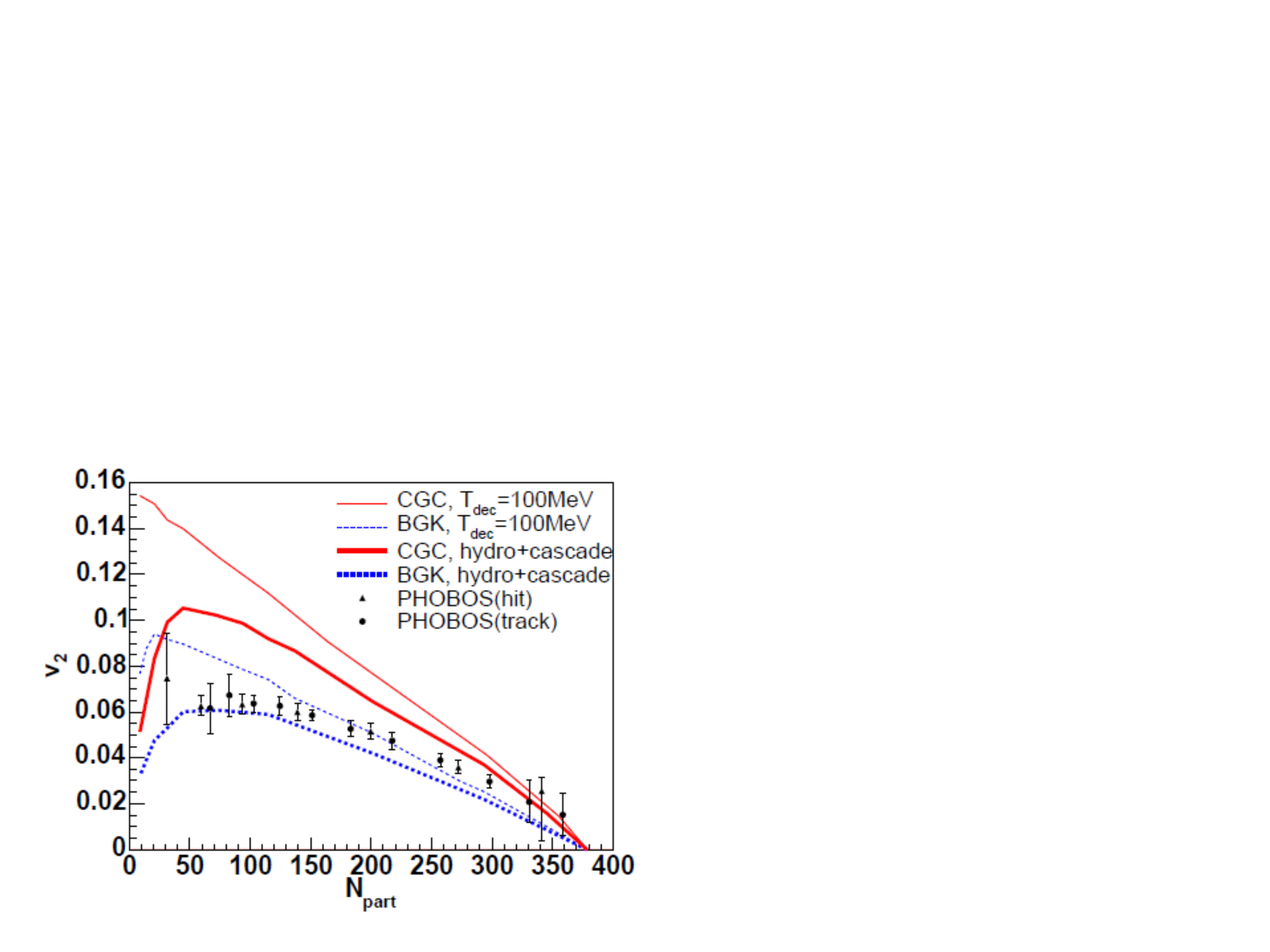}
  \caption{
Integrated elliptic flow parameter $v_{2}$ at midrapidity
as a function of the number of participants 
\cite{Hirano:2005xf}.
Red (Blue) solid line is the result from a hybrid model with CGC
 (Glauber) initial condition.
Dashed lines correspond to the results by assuming
kinetic freezeout happens at $T=100$ MeV.
\label{fig:v2cent}
}
\end{figure}

An announcement of discovery of perfect fluidity was made in 2005 \cite{bnl}.
This is based on a fact that
 elliptic flow parameters $v_{2}$ are reproduced 
remarkably well from 
ideal hydrodynamic calculations with Glauber type initial conditions 
\cite{Kolb:2000fh,Huovinen:2001cy,Kolb:2001qz,Hirano:2001eu,Teaney:2001av}.
Just after that, it was claimed ideal hydrodynamics with color glass condensate (CGC)
 initial condition does not reproduce
$v_{2}$ data \cite{Hirano:2005xf} as shown in Fig.~\ref{fig:v2cent}.
It is well known that $v_{2}$ is almost proportional to initial eccentricity
of the profile just after collisions.
So the discrepancy between the data and the model calculations
comes mainly from larger initial eccentricity from the CGC
model than that from the Glauber model.
For many years, understanding of initial conditions
in hydrodynamic models
had been very important.
Nevertheless, after this work, 
importance of discrimination of initial models was 
recognized again more than ever.
Nowadays 
it becomes a standard scheme to analyze the data
by comparison of  hydrodynamic results from several initial conditions
such as Glauber and CGC
with each other

Before $\sim$2007, most of hydro groups except for Rio de Janeiro-Sao Paulo group 
\cite{Osada:2001hw,Aguiar:2001ac,Hama:2004rr}
employed smooth initial conditions which could be identified with event-averaged
initial conditions.
However, event-by-event fluctuation in the initial conditions
turned out to be important 
after  the third order deformation parameter
resolves the ridge and Mach cone problems simultaneously \cite{Alver:2010gr}.
\begin{figure}[htbp]
  \centering
  \includegraphics[width=12cm, clip, bb=50 50 300 200]{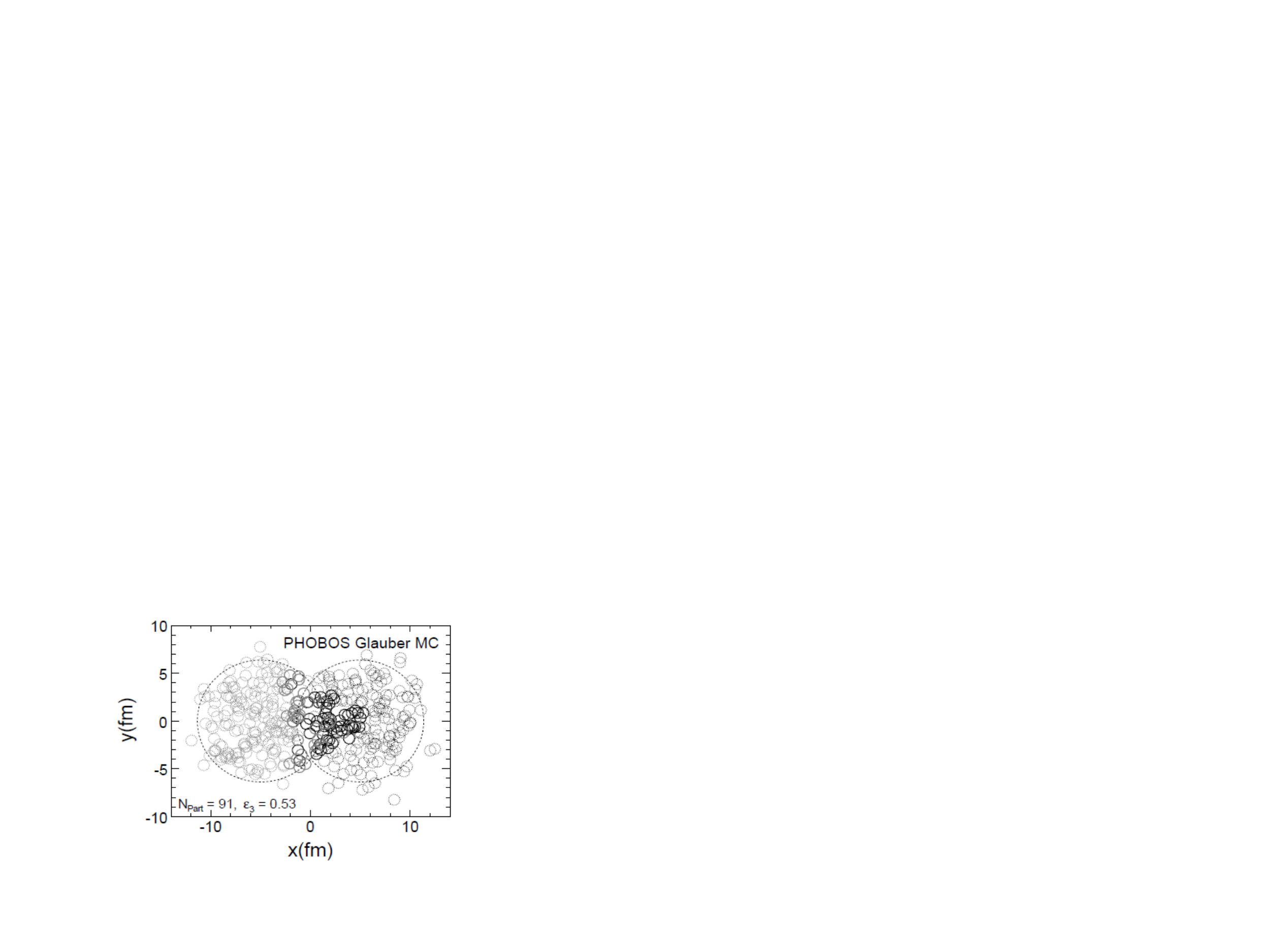}
  \caption{
An sample of event in the transverse plane from the PHOBOS Monte-Carlo Glauber model \cite{Alver:2010gr}.
\label{fig:phobosmc}
}
\end{figure}
Figure \ref{fig:phobosmc}
shows position of participants and spectators from the PHOBOS Monte-Carlo
Glauber model. In this particular event, a profile of
participants looks like a triangular shape.
The system responds to this initial profile and in particular in central events
the signal of triangular flow $v_{3}$ becomes comparable with that of elliptic flow.
Odd harmonics has never been considered seriously until then.

These two results opened up a new era of investigating
details of initial conditions and
triggered quite a lot of work on initial stages thereafter.
I think this is a part of the main reasons why
this conference was held.

Along the lines of the thought,  I highlight the topics of p/d-A collisions, isotropization,
thermalization, 
fluctuations and recent development in hydrodynamics and transport theory
in this review.

\section{p/d-A collisions}

One of the big surprises in the physics of high-energy nuclear collisions
is apparent collectivity of matter created in p/d-A collisions.
p/d-A collisions were called ``control experiment" 
to understand the so-called cold nuclear matter effects
such as Cronin effect, nuclear parton distribution and so on.
Basic consensus in the community was that no quark gluon plasma (QGP)
is created in such collisions due to its smallness.
However, recent experimental data show ridge structure
in p-p \cite{Khachatryan:2010gv} and p-A collisions \cite{CMS:2012qk,Abelev:2012ola}
 at LHC and finite $v_{2}$ in d-A collisions at RHIC \cite{Adare:2013piz}.

\begin{figure}[htbp]
  \centering
  \includegraphics[width=12cm, clip, bb=0 40 720 400]{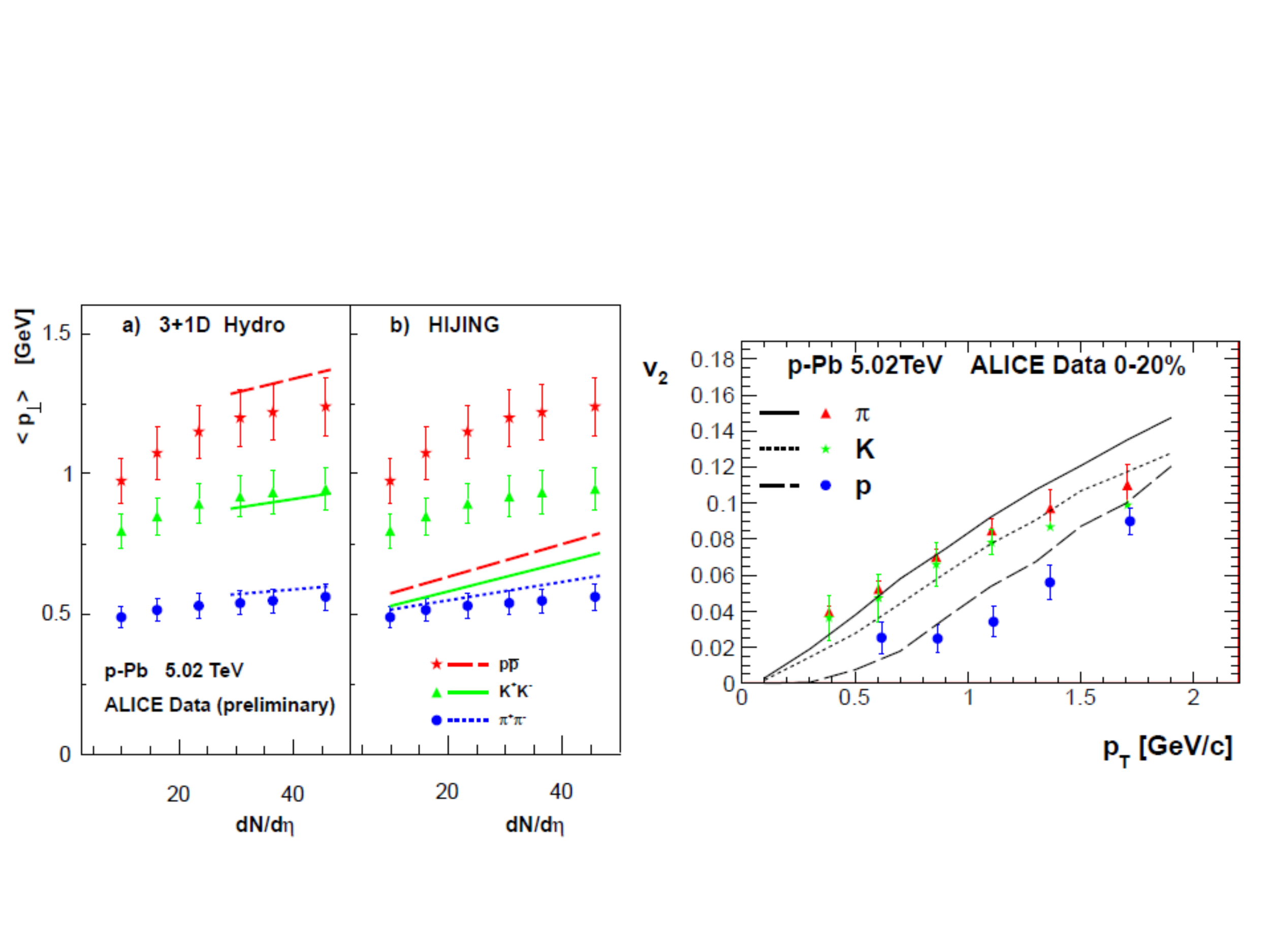}
  \caption{
Mean transverse momentum as a function of charged particle multiplicity
 (left) and differential elliptic flow parameter (right)
for particle identified hadrons in p+Pb collisions at 5.02 TeV \cite{Bozek:2013ska}.
\label{fig:bozek}
}
\end{figure}

The ridge structure in p-A collisions at the LHC energy
 can be understood 
as an initial state effect within the CGC picture qualitatively \cite{Dusling:2012wy}.
However, what is more surprising to us is 
mass ordering behavior, which has been
a strong signal of collectivity in A-A collisions,
is observed even in p-A collisions at LHC \cite{Abelev:2012ola}
and in d-A collisions at RHIC \cite{Adare:2013piz}.
Mean transverse momentum 
as a function of multiplicity
for particle identified
hadrons such as pions, kaons and protons
are well separated \cite{Chatrchyan:2013eya}, which indicates existence of radial flow.
Furthermore mass ordering pattern of $p_{T}$
differential elliptic flow $v_{2}(p_{T})$ \cite{ABELEV:2013wsa}
also indicates final rescatterings effects.
These observables are reasonably 
reproduced by employing a hydrodynamic model \cite{Bozek:2013ska}.
Figure \ref{fig:bozek} (left)
shows comparison of hydrodynamic results of mean $p_{T}$
 with experimental data.
It should be noted that HIJING, in which there is no rescatterings,
cannot reproduce this mass splitting pattern.
In  Fig.~\ref{fig:bozek} (right), ALICE $v_{2}(p_{T})$ data are compared with hydrodynamic results
and this hydrodynamic model reasonably describes the tendency of the data.

On the other hand, the color reconnection 
option in PYTHIA is discussed by Ortiz
and is found to result in apparent flow-like effect in particle ratio
as a function of $p_{T}$ \cite{Ortiz:2013yxa}.
It would be interesting to see  in this calculation whether the ridge-like structure 
also appears
in high-multiplicity p-p events.

One of the hydrodynamic results which I found intriguing in the conference  \cite{is2013}
is shock-wave pattern in d-A collisions shown by Schenke. 
\begin{figure}[htbp]
  \centering
  \includegraphics[width=12cm, clip, bb=0 150 720 400]{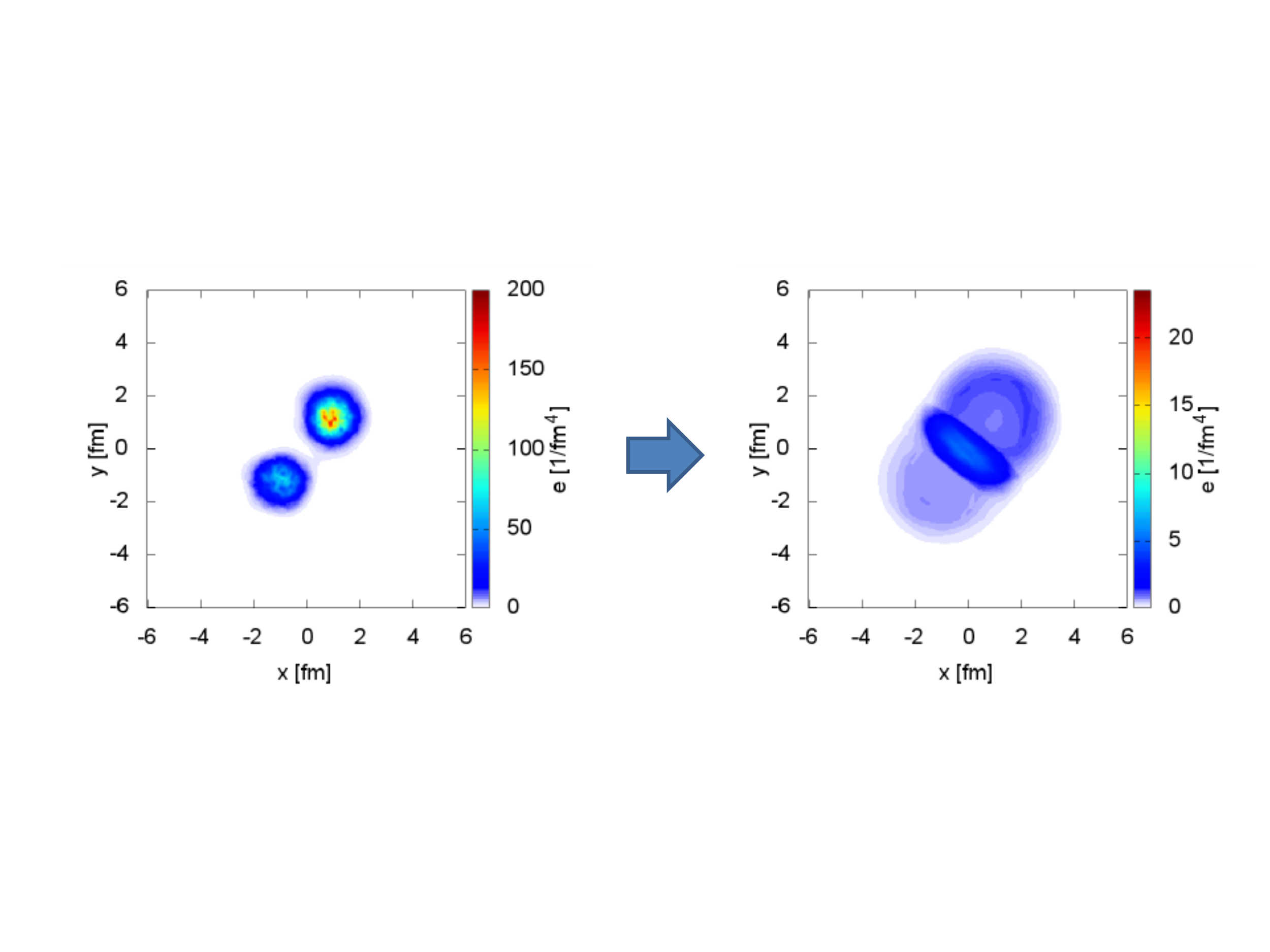}
  \caption{
Time evolution of energy density in d-Au collisions at the RHIC energy.
(Left) Initial condition at $\tau = 0.2$ fm/$c$. (Right) Energy density distribution at $\tau=5.2$ fm/$c$.
Figure is adapted from a talk by Schenke in this conference \cite{is2013}.
\label{fig:schenke}
}
\end{figure}
Figure~\ref{fig:schenke} shows time evolution of energy density in d-Au collisions at the RHIC energy.
This reminds us a volcano scenario by T.D.~Lee \cite{TDLee}.
``Squeeze-out" of matter can take place 
and substantial back-to-back correlation may appear
perpendicular to the axis of the two nucleons in the deuteron.
In this context, see also the pioneering work of hydrodynamic simulations
with bumpy initial conditions from HIJING
in A-A collisions \cite{Gyulassy:1996br}.

In any case, the good news is that the physics of high-energy
nuclear collisions has been more sophisticated.
Before RHIC started, most of the people in this community
did not believe hydrodynamic description of the QGP.
Just after RHIC launched,
hydrodynamic description immediately turned out to be
successful.
At that time, smooth initial conditions were employed,
which means that the size of coarse-graining was of the order of 5 fm.
In the last few years,
event-by-event initial fluctuation gets important
to understand higher order anisotropy.
The size of the fluctuation 
or, in turn, the size of the coarse-graining
is of the order of 1 fm or less.
Obviously, resolution to describe initial profile
is getting better.
Now there is a possibility for hydrodynamic framework
to work even in smaller system created in p-p or p/d-A collisions.

\section{Isotropization and thermalization}

Most of the people in this community agree that
a final piece of jig-saw puzzle
to solve high-energy nuclear collisions
is to understand how to thermalize the system just after collisions.
At leading order of CGC formalism, energy momentum tensor just after the first contact of
nuclear collision becomes \cite{Lappi:2006fp}
\begin{eqnarray}
T^{\mu \nu} (\tau=0^{+}) & = & \mathrm{diag}(\epsilon_{0}, \epsilon_{0}, \epsilon_{0}, -\epsilon_{0}),\\
\epsilon_{0} & = & \epsilon(\tau=0^{+}),
\end{eqnarray}
where $\epsilon$ is the energy density of the color fields. Note
that this energy momentum tensor is traceless due to scale invariance.
A remarkable feature is that negative pressure appear in the longitudinal direction.
This is something like an elastic body: Negative $pdV$ work stores the system with energy
through expansion of the system.
The question in high-energy nuclear collisions is
how to obtain the form of energy momentum tensor like
\begin{equation}
T^{\mu \nu} (\tau_{\mathrm{iso}/\mathrm{therm}}) = \mathrm{diag}(\epsilon(\tau_{\mathrm{iso}/\mathrm{therm}}), 
P_{T}, P_{T}, P_{L}),
\end{equation}
where $P_{T} \approx P_{L}$ and isotropization or thermalization time $\tau_{\mathrm{iso}/\mathrm{therm}}$ 
is of the order of 1 fm/$c$.

\begin{figure}[htbp]
  \centering
  \includegraphics[width=12cm, clip, bb=0 40 500 300]{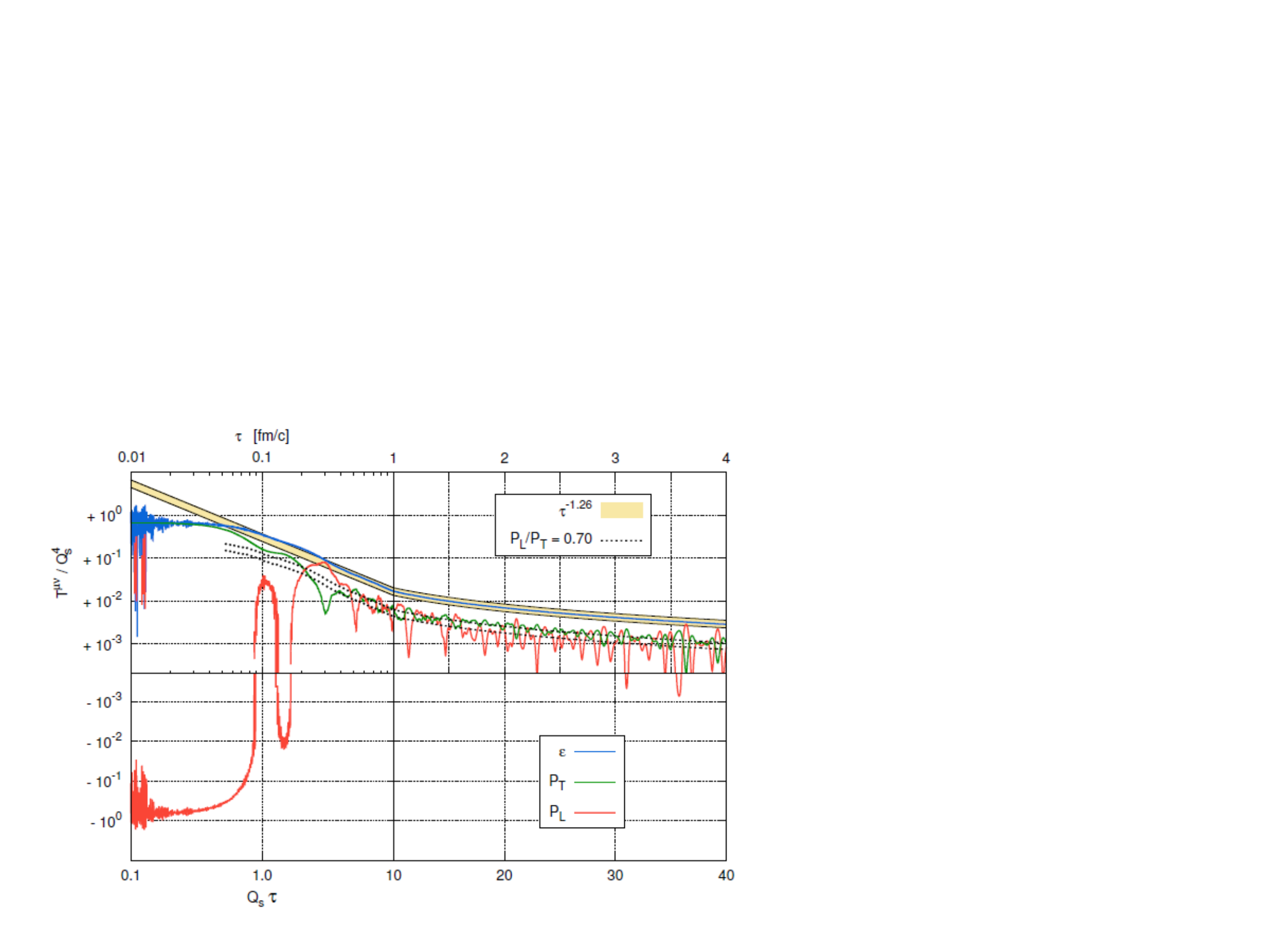}
  \caption{
Time evolution of longitudinal (red) and transverse (blue) pressure in an expanding coordinate.
This figure is adapted from a talk by T.~Epelbaum in the conference \cite{is2013}.
\label{fig:epelbaum}
}
\end{figure}

Temporal behavior of transverse and longitudinal
pressure is discussed by Epelbaum. Classical Yang-Mills equation
with the CGC initial conditions is solved in an expanding coordinate \cite{Gelis:2013rba}.
Figure \ref{fig:epelbaum} shows that
$P_{L} \sim 0.7 P_{T}$ at $\sim 0.4$ fm/$c$
and 
that the system
exhibits hydrodynamic behavior 
even for small coupling $\alpha_{s} \sim 10^{-2}$.

Anisotropic hydrodynamics helps us to describe
this stage before conventional
hydrodynamic regime in which isotropic pressure is required.
Details of the formalism and its consequences were discussed 
by Strickland in this conference \cite{Strickland:2014eua}.

\section{Fluctuation}

Topics of fluctuation in a broad sense
are popular for these years.
In this section, I discuss some aspects of
fluctuation in high-energy nuclear collisions.

\begin{figure}[htbp]
  \centering
  \includegraphics[width=12cm, clip, bb=0 40 300 200]{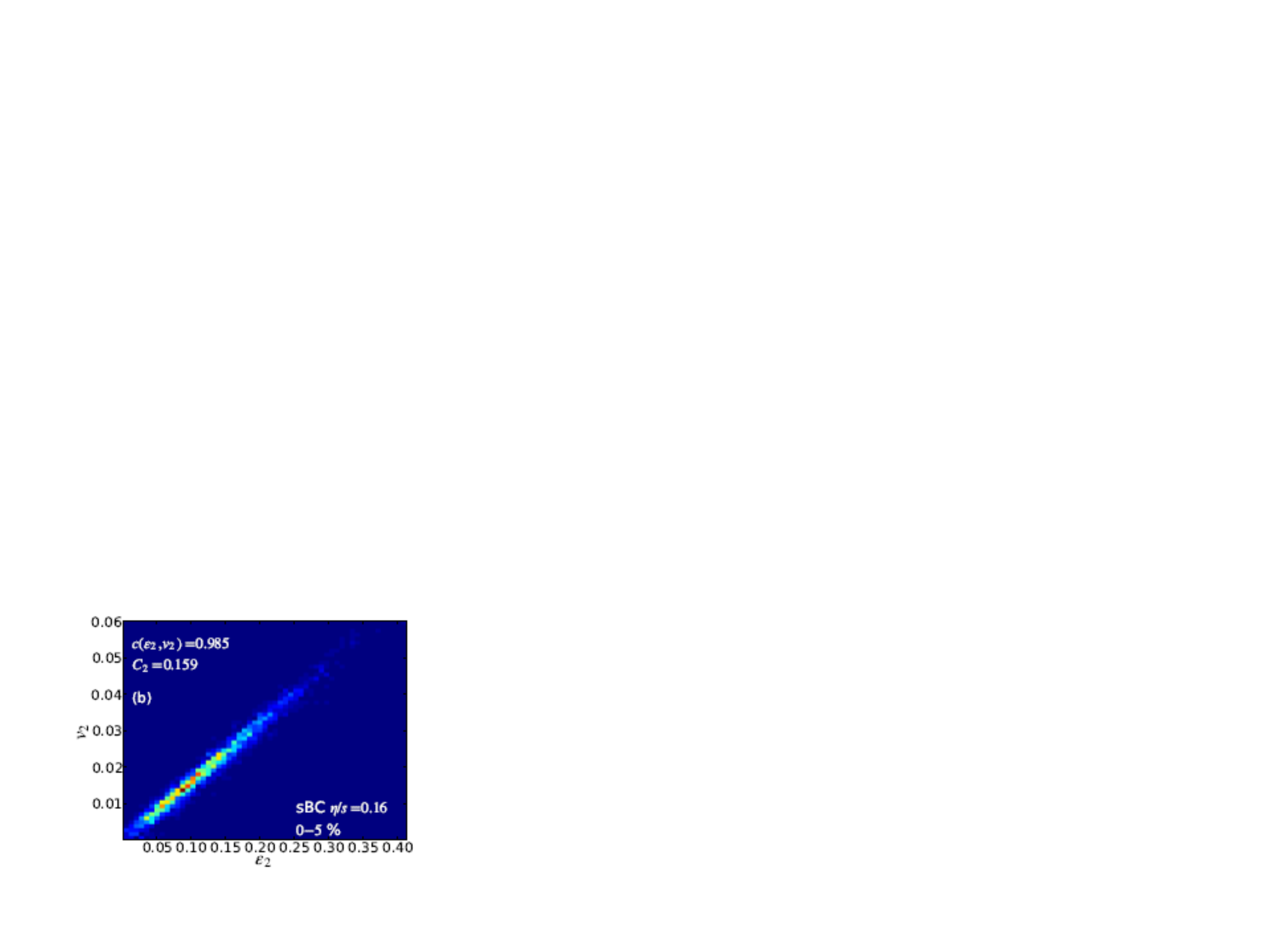}
  \caption{Correlation between initial eccentricity $\varepsilon_{2}$ and final elliptic flow parameter
$v_{2}$ from event-by-event viscous hydrodynamic simulations.
This figure is adapted from a talk by H.~Niemi \cite{is2013}.
\label{fig:niemi}
}
\end{figure}

Relation between
initial fluctuation of matter profile and final higher harmonics is the key to
investigate transport property of the system.
Figure \ref{fig:niemi}
shows strength of correlation between
initial eccentricity $\varepsilon_{2}$ and final elliptic flow parameter $v_{2}$
in viscous hydrodynamic calculations with the ratio of shear viscosity
to entropy density being $\eta/s = 0.16$ \cite{Niemi:2012aj}.
As is shown, final elliptic flow parameter is strongly correlated with initial eccentricity.
Regarding this, it would be interesting to see what happens to this correlation
if hydrodynamic fluctuation during evolution  is taken into account as discussed by Murase.
Constitutive equation in general can be written as a stochastic equation,
\begin{eqnarray}
\Pi(x) &=& \int d^{4}x' G_{R}(x,x')F(x') + \delta \Pi,\\
\langle \delta \Pi(x) \Pi(x') \rangle & = & T G^{*}(x, x').
\end{eqnarray}
Here $\Pi$ is the dissipative current, $G_{R}$ is the retarded Green function ($G^{*}$ being its 
symmetrized version with respect to time), $F$ is the thermodynamic force and
$\delta \Pi$ is the hydrodynamic fluctuation as a random force.
This is nothing but a fluctuation-dissipation relation. When dissipation
exists, fluctuation should appear in a consistent manner \cite{Murase:2013tma}.
These new sources of the fluctuation together with dissipative corrections
will be implemented in next-generation hydrodynamic simulations.

Retinskaya discussed an inverse problem by assuming the following
equation \cite{Retinskaya:2013gca}:
\begin{equation}
\label{eq:scalingharmonics}
v_{n} (\mathrm{exp. data}) = \left( \frac{v_{n} }{\varepsilon_{n}} \right)_{\mathrm{hydro}} \varepsilon_{n}.
\end{equation}
One can estimate $v_{n}/\varepsilon_{n}$ for a broad range
of $\varepsilon_{n}$ using viscous hydrodynamic simulations like Fig.~\ref{fig:niemi}.
Within these model calculations,
one can map experimetal $v_{n}$ data into higher order eccentricity $\varepsilon_{n}$
from Eq.~(\ref{eq:scalingharmonics}).
Thus a reasonable scaling relation $\varepsilon_{2}/\varepsilon_{3}^{k} = \mathrm{const.}$ is 
found with $k\sim 0.5$ for RHIC data and $k\sim 0.6$ for LHC data.
This result is obtained rather in a model-independent way in the initial stage.
So one can test whether one's favorite model for initial conditions
would obeys this scaling relation.
If not, the model could be discarded without performing massive hydrodynamic
simulations. 
For a detail of comparison among initial models, see Ref.~\cite{Retinskaya:2013gca}.

Conventionally hydrodynamic description is applicable when the spatial gradients of thermodynamic variables
are small enough.
This is one of the main reasons why hydrodynamic description
would not be trusted in small system such as p-p or p-A collisions.
Suppose that interaction region is large enough for hydrodynamics
to be applied even in p-p collisions due to fluctuation of inelastic cross section.
This is an idea of ``fat" proton advocated by Muller \cite{Coleman-Smith:2013rla}.
If the deposited energy is sufficient for thermalization and 
the gradients of thermodynamic variables is small enough,
there would be a chance for hydrodynamics to be applicable even in p-p collisions.

Fluctuation of saturation scale results in fluctuation of multiplicity.
In particular, high multiplicity p-p events can be reproduced by
IP-Sat model with fluctuating saturation scale \cite{Schenke:2013aza}.
Before going to A-A collisions, it is of particular importance to
understand mechanism of 
particle production in rather simpler system such as p-p collisions.

\section{Recent development in hydrodynamics and transport theory}

One of the good news is a revival of the final state saturation model 
discussed by Paatelainen.
This model is based on perturbative QCD parton production,
saturation of gluons in the pre-thermalization stage
and subsequent hydrodynamic evolution in (2+1)-dimensional space \cite{Paatelainen:2013eea}.
Just after RHIC started, one of the main observables
was centrality dependence of multiplicity, $(dN_{\mathrm{ch}}/d\eta)/(N_{\mathrm{part}}/2)$.
Several model predictions were
compared with the RHIC data and, in fact,
the final state saturation model \cite{Eskola:1999fc} did not do a good job
in this game \cite{PHOBOS}.
However, there was a misidentification of centrality between theoretical results with experimental data \cite{Kari}.
After correcting this, results from the final state saturation model
agree well with experimental data now.

Nowadays there are quite a lot of hydrodynamic simulations in the market.
Dissipative effects are taken into account in most of the models
directly by solving viscous hydrodynamic equations
and/or indirectly by combining hydrodynamic simulation with
subsequent kinetic evolution of hadron gases.
In this conference, we saw two brand-new viscous hydrodynamic
results from v-USPhydro \cite{Noronha-Hostler:2013gga} which is a successor of
NeXSPheRIO
and ECHO-QGP (Eulerian Conservative High-Order Code) \cite{DelZanna:2013eua}.
The main focus of v-USPhydro
is on the effect of  bulk viscosity which
has not been extensively discussed earlier.
Although ECHO was originally developed for astrophysics,
it is now applied to the physics of QGP.
Numerical tests are almost finished and now
they are going to analyze actual data using this code.

Denicol investigated the effect of non-linear terms with respect to
dissipative currents 
in the second order hydrodynamics equations \cite{Molnar:2013lta}
which have been missing so far in most of viscous hydrodynamic simulations.
As expected, the difference between with and without non-linear term
is manifested at large $\eta/s > 0.2$.
In the viscous hydro code in the next generation,
these non-linear terms should be taken into account.
 
Usually, it is almost impossible to incorporate critical behavior of
phase transition in kinetic theory.
This is the reason why hydrodynamics has an advantage against the kinetic theory.
In these years,
there is a trend that some of the hydrodynamic properties
are implemented in the kinetic approaches.
Marty discussed 
Nambu--Jona-Lasinio type phase transition in a kinetic theory \cite{Marty:2012vs}
 is combined with
the framework of 
PHSD (Parton-Hadron String Dynamics) \cite{Cassing:2009vt}.
Greco implemented a fixed $\eta/s$ in a transport model
and analyzed flow data \cite{Plumari:2013bga} to conclude evidence of phase transition.

\section{Summary}
p/d-A collisions provide us with a new opportunity to learn novel aspects of
 high energy hadron/nuclear reaction in a unified picture.
It is good to keep in touch with each other between sub-communities of initial stages (CGC, nuclear PDF, etc.)
 and final evolution (hydro, transport, etc.).
Future e-A program should shed light on
more precise structure of hadrons/nuclei at very high energy.

\section*{Acknowledgment}

I would like to thank C.~Salgado, N.~Armesto,  C.~Pajares
and other local organizers
for excellent organization and, in particular,
for providing us with a wonderful setting of
fruitful discussion in this conference.



\end{document}